\documentclass[12pt]{article}
\usepackage{amsmath,amssymb}
\usepackage{epsfig}
\usepackage{amssymb}

\usepackage{amssymb}
\usepackage{epsfig}

\def\a{\alpha}

\def\g{\gamma}

\def\d{\delta}

\def\beq{\begin{equation}}
\def\eeq{\end{equation}}
\def\beqn{\begin{eqnarray}}
\def\eeqn{\end{eqnarray}}
\def\ba{\begin{eqnarray}}
\def\ea{\end{eqnarray}}

\def\m{{\tt -}}

\def\xprim2bar{\overline{x}^{\prime\prime}}

\def\beq{\begin{equation}}
\def\eeq{\end{equation}}

\newcommand{\beqa}{\begin{eqnarray}}
\newcommand{\eeqa}{\end{eqnarray}}

\let\a=\alpha      \let\g=\gamma   \let\d=\delta
         
        \let\m=\mu

       \let\C=\Chi    
\let\a=\alpha      \let\g=\gamma   \let\d=\delta
         
        \let\m=\mu

       \let\C=\Chi    
\newcommand{\be}{\begin{equation}}
\newcommand{\ee}{\end{equation}}
\newcommand{\bea}{\begin{eqnarray}}
\newcommand{\eea}{\end{eqnarray}}

\usepackage{graphics}
\usepackage{graphicx}
\def\A5{(A_5)_{\rm lat}}
\def\thintablerule{\hrule height0.4pt}
\def\C{{\mathbb C}}

\def\one{\mbox{1 \kern-.59em {\rm l}}}

\def\bb#1{\mathbb {#1}}

\def\A#1{SU^{#1}(2)}

\begin{document}


\vskip 0.5cm
\centerline{\Large Reduction of ${\cal N}=1$, $E_8$ SYM over $SU(3)/U(1)\times U(1)\times \mathbb{Z}_3$}
\vskip .5cm
\centerline{\Large and its four-dimensional effective action}

\vskip 2 cm
\centerline{\large Nikos Irges$^1$ and George Zoupanos$^{2,1}$}
\vskip.5cm
\centerline{\it 1. Department of Physics}
\centerline{\it National Technical University of Athens}
\centerline{\it Zografou Campus, GR-15780 Athens Greece}
\vskip .4cm
\centerline{\it 2. Institut f\"ur Theoretische Physik}
\centerline{\it Universit\"at Heidelberg}
\centerline{\it Philosophenweg 16}
\centerline{\it D-69120 Heidelberg, Germany}
\vskip 1.5 true cm
\thintablerule
\vskip 2.0ex
\leftline{\bf Abstract}
We propose an extension of the Standard Model inspired by the 
$E_8\times E_8$ Heterotic String. In order that a reasonable effective Lagrangian is presented
we neglect everything else other than the ten-dimensional ${\cal N}=1$ supersymmetric Yang-Mills
sector associated with one of the gauge factors and certain couplings necessary for anomaly 
cancellation. 
We consider a compactified space-time 
$M_4 \times B_0/\mathbb{Z}_3$, where $B_0$ is the nearly K\"ahler manifold $SU(3)/U(1) \times U(1)$ and $\mathbb{Z}_3$ is a freely acting discrete group on $B_0$.
Then we reduce dimensionally the $E_8$ on this manifold and we employ the Wilson flux mechanism 
leading in four dimensions to an $SU(3)^3$ gauge theory with the spectrum of a ${\cal N}=1$ supersymmetric theory.
We compute the effective four-dimensional Lagrangian and demonstrate that an
extension of the Standard Model is obtained with interesting features including a
conserved baryon number and fixed tree level Yukawa couplings and scalar potential.
The spectrum contains new states such as right handed neutrinos and heavy vector-like quarks.

\vskip 1.0ex\noindent
\vskip 2.0ex
\thintablerule

\vskip-0.2cm
\newpage

\section{Introduction}

Building extensions of the Standard Model and of its supersymmetric version with a solid motivation 
from fundamental unified theories is an important program, one of the reasons being that a typical model 
of this type is more constrained than its generic counterpart. 
We concentrate here on the class of models obtained from the compactification of the
10-dimensional supersymmetric ${\cal N}=1$, $E_8$ Yang-Mills gauge theory on coset spaces
of Lie groups. We refer to such dimensional reduction as Coset Space Dimensional Reduction (CSDR).
More specifically, we consider the six-dimensional nearly K\"ahler manifold $S/R = SU(3)/U(1) \times U(1)$ moded out 
by a freely acting $\mathbb{Z}_3$, as the manifold of the extra dimensions and we show that,
in the presence of a certain $\mathbb{Z}_3$ Wilson Loop (WL), it yields an interesting constrained, softly broken supersymmetric
extension of the Standard Model (SM). The terms that break supersymmetry are in fact soft provided that
the supersymmetry breaking scale and the compactification scale are both in the few TeV regime \cite{ignatios}. 

We start with a short review of the CSDR construction.
For a detailed exposition see \cite{Review}. 
Consider a Lie group $S$ and its subgroup $R$.
These define a $d$-dimensional coset $S/R$ on which the extra dimensions 
of $M^{4}\times S/R$ are compactified ($M^4$ is our space-time).
$S$ acts as a symmetry group on the extra
coordinates. The CSDR scheme demands that an $S$-transformation
of the extra $d$ coordinates is a gauge transformation of the
fields that are defined on $M^{4}\times S/R$,  thus a gauge
invariant Lagrangian written on this space is independent of the
extra coordinates. Fields defined in this way are called symmetric.
Consider further a $D$-dimensional
Yang-Mills-Dirac theory with gauge group $G$ defined on a
manifold $M^{D}$ compactified on $M^{4}\times S/R$, $D=4+d$, $d=dimS-dimR$:
\begin{equation}
A=\int d^{4}xd^{d}y\sqrt{-g}\Bigl[-\frac{1}{4}
Tr\left(F_{MN}F_{K\Lambda}\right)g^{MK}g^{N\Lambda}
+\frac{i}{2}\overline{\psi}\Gamma^{M}D_{M}\psi\Bigr] ,
\end{equation}
where $D_{M}= \partial_{M}-\theta_{M}-A_{M}$, 
with $\theta_{M}=\frac{1}{2}\theta_{MN\Lambda}\Sigma^{N\Lambda}$ the spin connection of $M^{D}$, and
$F_{MN}
=\partial_{M}A_{N}-\partial_{N}A_{M}-\left[A_{M},A_{N}\right]$,
where $M$, $N$ run over the $D$-dimensional space. The fields
$A_{M}$ and $\psi$ are symmetric.  
The only constraint on the fermion representation $F$ of $G$ comes from supersymmetry, in case it is imposed.
Let $\xi_{A}^{\alpha}$, $A=1,\ldots,dimS$, be the Killing vectors which generate the
symmetries of $S/R$ and $\Omega_{A}$ the compensating gauge
transformation associated with $\xi_{A}$. Defining the
infinitesimal coordinate transformation as $\delta_{A} \equiv
L_{\xi_{A}}$, the Lie derivative with respect to $\xi$, we
have for the scalar,vector and spinor fields:
\begin{eqnarray}
\delta_{A}\phi&=&\xi_{A}^{\alpha}\partial_{\alpha}\phi=D(\Omega_{A})\phi,
\nonumber \\
\delta_{A}A_{\alpha}&=&\xi_{A}^{\beta}\partial_{\beta}A_{\alpha}+\partial_{\alpha}
\xi_{A}^{\beta}A_{\beta}=\partial_{\alpha}\Omega_{A}-[\Omega_{A},A_{\alpha}],\nonumber\\
\delta_{A}\psi&=&\xi_{A}^{\alpha}\psi-\frac{1}{2}G_{Abc}\Sigma^{bc}\psi=
D(\Omega_{A})\psi. \label{constr}
\end{eqnarray}
$\Omega_{A}$ depend only on internal coordinates $y$ and $D(\Omega_{A})$
represents a gauge transformation in the appropriate
representation of the fields. $G_{Abc}$ represents a tangent space
rotation of the spinor fields. The variations $\delta_{A}$
satisfy, $[\delta_{A},\delta_{B}]=f_{AB}^{\\C}\delta_{C}$ and
lead to the consistency relation for $\Omega_{A}$'s
$\xi_{A}^{\alpha}\partial_{\alpha}\Omega_{B}-\xi_{B}^{\alpha}\partial_{\alpha}
\Omega_{A}-\left[\Omega_{A},\Omega_{B}\right]=f_{AB}^{\ \ C}\Omega_{C}$.
The $\Omega$'s transform under a gauge transformation as
$\widetilde{\Omega}_{A} = g\Omega_{A}g^{-1}+(\delta_{A}g)g^{-1}$.
Using the fact that the Lagrangian is independent of
$y$ we can do all calculations at $y=0$ and choose a gauge where $\Omega_{a}=0$.

The analysis of the constraints eq. (\ref{constr}) 
\cite{Review,Manton} provides us with the four-dimensional
unconstrained fields as well as with the gauge invariance that
remains in the theory after dimensional reduction.
The components $A_{\mu}(x,y)$ of the initial gauge
field $A_{M}(x,y)$ become, after dimensional reduction, the
four-dimensional gauge fields and furthermore they are independent
of $y$. In addition one can find that they have to commute with
the elements of the $R_{G}$ subgroup of $G$. Thus the
four-dimensional gauge group $H$ is the centralizer of $R$ in $G$,
$H=C_{G}(R_{G})$. Similarly, the $A_{\alpha}(x,y)$ components of
$A_{M}(x,y)$ denoted by $\phi_{\alpha}(x,y)$ from now on, become
scalars at four dimensions. These fields transform under $R$ as a
vector $v$, i.e.
\begin{eqnarray}
S &\supset& R \nonumber \\
adjS &=& adjR+v.
\end{eqnarray}
Moreover $\phi_{\alpha}(x,y)$ act as an intertwining operator
connecting induced representations of $R$ acting on $G$ and $S/R$.
This implies, exploiting Schur's lemma, that the transformation
properties of the fields $\phi_{\alpha}(x,y)$ under $H$ can be
found if we express the adjoint representation of $G$ in terms of
$R_{G} \times H$ :
\begin{eqnarray}
G &\supset& R_{G} \times H \nonumber \\
 adjG &=&(adjR,1)+(1,adjH)+\sum(r_{i},h_{i}).
\end{eqnarray}
Then if $v=\sum s_{i}$, where each $s_{i}$ is an irreducible
representation of $R$, there survives an $h_{i}$ multiplet for
every pair $(r_{i},s_{i})$, where $r_{i}$ and $s_{i}$ are
identical irreducible representations of $R$.

The fermions,
\cite{Review,Slansky,Chapline,Palla} just as the scalars,
act as intertwining operators between induced representations
acting on $G$ and the tangent space of $S/R$, $SO(d)$.  
The representation $F$ of the
initial gauge group in which the fermions transform decompose under
$R_{G} \times H$ as
\begin{equation}
F= \sum (t_{i},h_{i}),
\end{equation}
and the spinor of $SO(d)$ under $R$ as
\begin{equation}
\sigma_{d} = \sum \sigma_{j}.
\end{equation}
Then for each pair $t_{i}$ and $\sigma_{i}$, where $t_{i}$ and
$\sigma_{i}$ are identical irreducible representations there is an
$h_{i}$ multiplet of spinor fields in the four-dimensional theory.
Specifying now the discussion to $D=10$, $d=6$  and considering an ${\cal N}=1$ 10-dimensional gauge theory which requires 
that fermions belong in the adjoint representation,
in order to obtain chiral fermions in the 4-dimensional effective
theory the Majorana and Weyl conditions have to be imposed in the higher dimensional theory.

\section{The low energy effective action}

In this section we summarise a few known facts about the dimensional reduction of the ${\cal N} = 1$, $E_8$ SYM over $SU(3)/U(1) \times U(1)$ \cite{CSDR}.
To determine the four-dimensional gauge group, the embedding of $R = U(1) \times U(1)$ in $E_8$ is suggested by the decomposition
\be
E_8 \supset E_6\times SU(3) \supset E_6 \times U(1)_A \times U(1)_B.
\ee
Then according to the rules discussed in sect. 1, the four-dimensional gauge group after dimensional reduction of $E_8$ under $SU(3)/U(1) \times U(1)$ is
\be
H = C_{E_8} (U(1)_A \times U(1)_B) = E_6 \times U(1)_A \times U(1)_B.
\ee
In order to determine the surviving scalars and fermions in four dimensions one further needs the explicit decomposition of the adjoint 
representation of $E_8$, 248 under $U(1)_A \times U(1)_B$ given by
\begin{eqnarray}
 248 = 1_{(0,0)}+1_{(0,0)}+1_{(3,\frac{1}{2})}+1_{(-3,\frac{1}{2})}+
1_{(0,-1)}+1_{(0,1)}+1_{(-3,-\frac{1}{2})}+1_{(3,-\frac{1}{2})}+\nonumber\\
78_{(0,0)}+27_{(3,\frac{1}{2})}+27_{(-3,\frac{1}{2})}+27_{(0,-1)}+
\overline{27}_{(-3,-\frac{1}{2})}+\overline{27}_{(3,-\frac{1}{2})}
+\overline{27}_{(0,1)}.\label{E8}
\end{eqnarray}
The $R=U(1) \times U(1)$
content of $SU(3)/U(1) \times U(1)$ vector and spinor are $$(3,\frac{1}{2})+(-3,\frac{1}{2})
+(0,-1)+(-3,-\frac{1}{2})+(3,-\frac{1}{2})+(0,1)$$ and
$$(0,0)+(3,\frac{1}{2})+(-3,\frac{1}{2}) +(0,-1)$$ respectively.
Applying the CSDR rules one finds that the surviving fields in
four dimensions are three ${\cal N}=1$ vector supermultiplets
$U^{\alpha},U_A,U_B$, (where $\alpha$ is an $E_{6}$, $78$
index and the other two refer to the two $U(1)'s$) containing the
gauge fields of $E_6\times U(1)_A \times U(1)_B$. The matter
content consists of three ${\cal N}=1$ chiral multiplets ($A^{i}$,
$B^{i}$, $C^{i}$) with $i$ an $E_{6}$, $27$ index and three ${\cal
N}=1$ chiral multiplets ($A$, $B$, $C$) which are $E_{6}$ singlets
and carry $U(1)_A \times U(1)_B$ charges.

To determine the potential one examines the decomposition
of the adjoint of the specific $S=SU(3)$ under $R=U(1) \times
U(1)$, i.e.
$SU(3) \supset U(1) \times U(1)$:
\begin{eqnarray}
 8 = (0,0)+(0,0)+(3,\frac{1}{2})+(-3,\frac{1}{2})
+(0,-1)+(-3,-\frac{1}{2})+(3,-\frac{1}{2})+(0,1).
\end{eqnarray}
Then the generators of $SU(3)$
can be accordingly grouped as
\begin{equation}
Q_{SU(3)} = \{Q_{0},Q'_{0},Q_{1},Q_{2},Q_{3},Q^{1},Q^{2},Q^{3} \}.
\end{equation}
This decomposition suggests the following change in  the notation of the scalar fields appearing in eq. (\ref{E8}),
\begin{equation}
(\phi_{I}, I=1,\ldots,8) \longrightarrow ( \phi_{0}, \phi'_{0},
\phi_{1}, \phi^{1}, \phi_{2}, \phi^{2}, \phi_{3}, \phi^{3}).
\end{equation}

In order to determine the potential in terms of the unconstrained fields we have to consider the 
commutation relations (CR's) of the generators of $E_8$, grouped also according to the decomposition eq. (\ref{E8}) as
\be
Q_{E_{8}}=\{Q_{0},Q'_{0},Q_{1},Q_{2},Q_{3},Q^{1},Q^{2},Q^{3},Q^{\alpha},
Q_{1i},Q_{2i},Q_{3i},Q^{1i},Q^{2i},Q^{3i} \},\label{E8group}
\ee
where, $ \alpha=1,\ldots,78 $ and $ i=1,\ldots,27 $.  
Examining the CR's of the generators  of $E_8$ grouped as in eq. (\ref{E8group}) we find that the redefined fields are subject to the constraints
\begin{eqnarray}
&& \left[\phi_{1},\phi_{0}\right]=\sqrt{3}\phi_{1}, \,\,\,\,\,\,\,\,\, \left[\phi_{3},\phi_{0}\right]=0,\,\,\,\,\,\,\,\,\,
\left[\phi_{1},\phi_{0}'\right]=\phi_{1} \nonumber \\
&& \left[\phi_{2},\phi_{0}\right]=-\sqrt{3}\phi_{2},  \,\,\,\,\,
\left[\phi_{2},\phi_{0}'\right]=\phi_{2}, \,\,\,\,\,\, \left[\phi_{3},\phi_{0}'\right]=-2\phi_{3} \, .
\end{eqnarray}
The solutions to the constraints in terms of the genuine
Higgs fields and the $E_{8}$ generators corresponding to
the embedding of $R=U(1) \times U(1)$ in  the $E_{8}$ are,
$\phi_{0}=\Lambda Q_{0}$ and $\phi_{0}'=\Lambda' Q_{0}'$,with
$\Lambda=\Lambda'=\frac{1}{\sqrt{10}}$, and
\be
\phi_{1} = R_{1} \alpha^{i} Q_{1i}+R_{1} \alpha Q_{1}, \;\;\;\;
\phi_{2} = R_{2} \beta^{i} Q_{2i}+ R_{2} \beta Q_{2}, \;\;\;\;
\phi_{3} = R_{3} \gamma^{i} Q_{3i}+ R_{3} \gamma Q_{3},
\ee
where the unconstrained  scalar fields transform under $E_{6}\times U(1)_A\times U(1)_B$ as
\be
\alpha_{i} \sim 27_{(3,\frac{1}{2})}, \;\;\; \beta_{i} \sim 27_{(-3,\frac{1}{2})}, \;\;\; \gamma_{i} \sim 27_{(0,-1)}, \;\;\;
\alpha \sim 1_{(3,\frac{1}{2})},\;\;\; \beta \sim 1_{(-3,\frac{1}{2})}, \;\;\;
\gamma \sim 1_{(0,-1)}
\ee
and $R_{1},R_{2},R_{3}$ are the coset space radii.
The scalar potential in terms of the independent fields can be then straightforwardly derived
and it is positive definite.
It turns out that the terms in the effective action can be arranged in 
a softly broken supersymmetric form.
In fact, the $F$-terms are obtained from the superpotential
\begin{equation}
{ \cal W }(A^{i},B^{j},C^{k},A,B,C)
=\sqrt{40}d_{ijk}A^{i}B^{j}C^{k} + \sqrt{40}ABC.
\end{equation}
$d$ is the fully symmetric $E_6$ invariant tensor.
The $D$-terms are $$D^{\alpha}= \frac{1}{\sqrt{3}}
\biggl(\alpha^{i}(G^{\alpha})_{i}^{j}\alpha_{j}
+\beta^{i}(G^{\alpha})_{i}^{j}\beta_{j}
+\gamma^{i}(G^{\alpha})_{i}^{j}\gamma_{j}\biggr),$$ $$D_{1}=
\sqrt{ \frac{10}{3} }\biggl(\alpha^{i}(3\delta_{i}^{j})\alpha_{j}
+ \overline{\alpha}(3)\alpha +
\beta^{i}(-3\delta_{i}^{j})\beta_{j} + \overline{\beta}(-3)\beta
\biggr),$$ $$D_{2} = \sqrt{ \frac{40}{3}
}\biggl(\alpha^{i}(\frac{1}{2}\delta_{i}^{j})\alpha_{j} +
\overline{\alpha}(\frac{1}{2})\alpha +
\beta^{i}(\frac{1}{2}\delta^{j}_{i})\beta_{j} +
\overline{\beta}(\frac{1}{2})\beta +
\gamma^{i}(-1\delta_{i}^{j})\gamma_{j} +
\overline{\gamma}(-1)\gamma \biggr)$$ and correspond to the
$E_{6} \times U(1)_A \times U(1)_B$ structure of the gauge
group. The rest of the terms are the trilinear and mass terms (specified below), which break
supersymmetry softly.
Finally the gaugino obtains a mass of the order ${\cal O}(R^{-1})$, which can be further adjusted by the torsion parameter.
\footnote{For a definition of the torsion parameter and how it enters this construction see \cite{CSDR, MZ}.}

\section{Wilson flux breaking}

Clearly, we need to further reduce the gauge symmetry. We will employ the Wilson flux breaking mechanism.
Let us briefly recall the Wilson flux mechanism for breaking spontaneously a gauge theory. 

\subsection{The Wilson flux mechanism}

Instead of considering a gauge theory on $M^{4}\times B_{0}$, with $B_{0}$ a simply connected manifold, and in our case a
coset space $B_{0}=S/R$, we consider a gauge theory on $M^{4}\times B$, with $B=B_{0}/F^{S/R}$ and $F^{S/R}$ a
freely acting discrete symmetry of $B_{0}$. It turns out that $B$ becomes
multiply  connected, which means that there will be contours not contractible to a point due to holes in the
manifold. For each element $g\in F^{S/R}$, we  pick up an element $U_{g}$ in $H$, i.e. in the four-dimensional
gauge group of the reduced theory, which can be represented as the Wilson loop
\begin{equation}
U_{g}=\mathcal{P}exp\left(-i~\int_{\gamma_g}T^{a}A_{M}^{a}(x)dx^{M}\right)\,,
\end{equation}
where $A_{M}^{a}(x)$ are vacuum $H$ fields with group generators $T^{a}$, $\gamma_g$ is a contour representing
the abstract element $g$ of $F^{S/R}$, and $\mathcal{P}$ denotes the path ordering.
Now if $\gamma_g$ is chosen not to be contractible to a point, then $U_g\neq 1$ although the vacuum field 
strength vanishes everywhere. In this  way an homomorphism of $F^{S/R}$ into $H$ is induced with image
$T^{H}$, which is the subgroup of $H$ generated by $\{U_g\}$. A field $f(x)$ on  $B_0$ is obviously equivalent
to another field on $B_0$ which obeys $f(g(x))=f(x)$ for every $g\in F^{S/R}$. However in the presence of the
gauge group $H$ this statement can be generalized to
\begin{equation}
\label{eq:Wilson-symmetry}
f(g(x))=U_{g}f(x)\,.
\end{equation}
The discrete symmetries $F^{S/R}$, which act freely on coset spaces $B_0=S/R$ are the center of $S$,
$\mathrm{Z}(S)$ and $\mathrm{W}=\mathrm{W}_{S}/\mathrm{W}_{R}$, where $\mathrm{W}_{S}$ and $\mathrm{W}_{R}$
are the Weyl groups of $S$ and $R$, respectively.  
The case of our interest here is
\beq
F^{S/R}= \bb{Z}_{3}\subseteq {\rm W}\, .
\eeq

\subsection{$SU(3)^3$ due to Wilson flux}

In order to derive the projected theory in the presence of the WL, one has to keep the
fields which are invariant under the combined action of the discrete group $\mathbb{Z}_3$ on the geometry and on the gauge indices.
The discrete symmetry acts non-trivially on the gauge fields and on the matter in the $27$ and the singlets. 
The action on the gauge indices is implemented via the matrix \cite{fuzzy}
${\rm diag}({\bf 1}_9,  \omega {\bf 1}_9,  \omega^2 {\bf 1}_9)$
with $\omega=e^{2i\pi/3}$. Thus, the gauge fields that survive the projection are those that satisfy
\be
A_\m = \g_3 A_\m \g_3^{-1}\, ,\label{Aproj}
\ee
while the surviving components of the matter fields in the 27's are those that satisfy
\be
\vec \a = \omega \g_3 \vec\a\, , \hskip .5cm \vec\beta = \omega^2 \g_3 \vec\beta\, , \hskip .5cm \vec\g = \omega^3 \g_3 \vec\g\, .
\label{matterproj}
\ee
Finally, the projection on the complex scalar singlets is
\be
\a = \omega \a\, , \hskip .5cm \beta = \omega^2 \beta\, , \hskip .5cm \g = \omega^3 \g\, .
\ee
It is easy to see then that after the $\mathbb{Z}_3$ projection the gauge group reduces to
\be
A_\m^A,\hskip 1cm A\in SU(3)_c\times SU(3)_L\times SU(3)_R
\ee
(the first of the $SU(3)$ factors is the SM colour group)
and the scalar matter fields are in the bi-fundamental representations
\be
{H_1} \sim ({\bar 3},1,{3})_{(3,1/2)}, \hskip .5cm 
{H_2} \sim ({3},{\bar 3},1)_{(0,-1)}, \hskip .5cm
{H_3} \sim (1,{3},{\bar 3})_{(-3,1/2)}.
\ee
There are also fermions in similar representations. 
Clearly, the Higgs is identified with the 9-component vector ${H_3}_a, a=1,\cdots ,9$.
Among the singlets, only ${\g}_{(0,-1)}$ survives.
In the following we will be using indices $a,b,c\cdots$ to count the complex components of a given
bi-fundamental representation and $i,j,k,\cdots = 1,2,3$ the different bifundamental representations.
 
Before we write the explicit scalar potential, we 
take appropriate actions such that there are 3 identical flavours from each of
the bifundamental fields. This can, in general, be achieved by introducing non-trivial windings in $R$.
We denote the resulting three copies of the bifundamental fields as (we will be using the index $l=1,2,3$ to specify the flavours)
\bea
&& 3\cdot {H_1} \longrightarrow  {H_1}^{(l)} \sim 3\cdot  ({\bar 3},1,{3})_{(3,1/2)}  \nonumber\\
&& 3\cdot {H_2} \longrightarrow  {H_2}^{(l)}  \sim 3\cdot ({3},{\bar 3},1)_{(0,-1)}  \nonumber\\
&& 3\cdot {H_3} \longrightarrow  {H_3}^{(l)} \sim 3\cdot (1,{3},{\bar 3})_{(-3,1/2)}\, .
\eea
Similarly we denote the three copies of the scalar as
\be
3\cdot {\g}_{{(0,-1)}}  \longrightarrow \theta^{(l)}_{{(0,-1)}}\, . 
\ee
The scalar potential gets accordingly three copies of each contribution.

In the following when it does not cause confusion 
we denote a chiral superfield and its scalar component with the same letter. Also, it is clear that the potential after the 
projection will have the same form as before the projection
with the only difference that only $\theta^{(l)}$ is non-vanishing among the singlets and that the sums over 
components now run only over the even under the projection components.

We can now rewrite the scalar potential as
\be
V_{\rm sc} = 3(3\Lambda^{2}+\Lambda'^{2})\biggl(\frac{1}{R_{1}^{4}}+\frac{1}{R_{2}^{4}}\biggr)
+\frac{3\cdot 4\Lambda'^{2}}{R_{3}^{4}} + \sum_{l=1,2,3} V^{(l)}
\ee
where
\be
V^{(l)} = V_{\rm susy} + V_{\rm soft}
\ee
with $V_{\rm susy} = V_D + V_F$.
Since there are three identical contributions to the potential, at least until we give vevs to the Higgses
(which in general can be different for each $l$) we can drop the flavour superscript $(l)$ from most of the fields.
Then, the explicit form of the $D$ and $F$ terms are
\bea
V_D &=& \frac{1}{2}\sum_A D^AD^A + \frac{1}{2}D_1D_1 + \frac{1}{2}D_2D_2 \nonumber\\
V_F &=&  \sum_{i=1,2,3} |F_{{H}_i}|^2 + |F_{\theta}|^2\, , \hskip .5cm F_{{H_i}} = \frac{\partial {\cal W}}{\partial {H_i}} ,  \hskip .5cm F_{{\theta}} = \frac{\partial {\cal W}}{\partial {\theta}}\, .
\eea
The $F$-terms derive from
\be
{\cal W} = \sqrt{40} d_{abc} H_1^aH_2^bH_3^c
\ee
and the $D$-terms are 
\bea
D^A &=& \frac{1}{\sqrt{3}} \langle H_i | G^A | H_i \rangle \nonumber\\
D_1 &=& 3\sqrt{\frac{10}{3}}\left( \langle H_1 | H_1 \rangle -  \langle H_2 | H_2 \rangle \right)\nonumber\\
D_2 &=& \sqrt{\frac{10}{3}} \left( \langle H_1 | H_1 \rangle +  \langle H_2 | H_2 \rangle -2  \langle H_3 | H_3 \rangle - 2|\theta |^2 \right)\, ,
\eea
where 
\bea
\langle H_i | G^A | H_i \rangle &=& \sum_{i=1,2,3}H_i^a (G^A)^b_a{H_i}_b \nonumber\\
\langle H_i | H_i \rangle &=&  \sum_{i=1,2,3} H_i^a\d^b_a {H_i}_b\, .
\eea
Finally the soft breaking terms are
\bea
V_{\rm soft} &=& \left(\frac{4R_1^2}{R_2^2R_3^2}-\frac{8}{R_1^2}\right) \langle H_1 | H_1 \rangle + \left(\frac{4R_2^2}{R_1^2R_3^2}-\frac{8}{R_2^2}\right) \langle H_2 | H_2 \rangle \nonumber\\
&+& \left(\frac{4R_3^2}{R_1^2R_2^2}-\frac{8}{R_3^2}\right) (\langle H_3 | H_3 \rangle +  |\theta |^2) \nonumber\\
&+& 80\sqrt{2} \left(\frac{R_1}{R_2R_3}+\frac{R_2}{R_1R_3} + \frac{R_3}{R_1R_2}\right)
(d_{abc} H_1^aH_2^bH_3^c + {\rm h.c.}).
\eea
The $(G^A)^b_a$ are structure constants, thus antisymmetric in $a$ and $b$.
The vector $|\phi \rangle$ and its hermitian conjugate $\langle \phi |$ represent the 9-dimensional bi-fundamental fields shown above.

The potential can be written in a more convenient form, as suggested in \cite{kephart}. 
It amounts to writing the vectors in complex $3\times 3$ matrix notation. 
The various terms in the scalar potential can be then interpreted as invariant Lie algebra polynomials.
We identify
\be
{H_1} \sim ({\bar 3},1,3) \longrightarrow N^\a_p\hskip .5cm
{H_2} \sim (3,{\bar 3},1) \longrightarrow L^a_\a \hskip .5cm
{H_3} \sim (1,3,{\bar 3}) \longrightarrow M^p_a
\ee
and introduce
\be
{\hat N}_\a^p = \frac{1}{3} \frac{\partial I_3}{\partial N^\a_p}\hskip .5cm
{\hat M}_p^a = \frac{1}{3} \frac{\partial I_3}{\partial M^p_a}\hskip .5cm
{\hat L}_a^\a = \frac{1}{3} \frac{\partial I_3}{\partial L^a_\a}\, ,
\ee
where
\be
I_3 = {\rm det N} + {\rm det M} + {\rm det L} -{\rm tr}(NML)\, .
\ee
In terms of these matrices, we have
$\langle H_1|H_1\rangle = {\rm tr} (N^\dagger N)$, 
$\langle H_2|H_2\rangle = {\rm tr} (L^\dagger L)$,
$\langle H_3|H_3\rangle = {\rm tr} (M^\dagger M)$
and 
\be
d_{abc} H_1^aH_2^bH_3^c = {\rm det N^\dagger} + {\rm det M^\dagger} + {\rm det L^\dagger} -{\rm tr}(N^\dagger M^\dagger L^\dagger)\, .
\ee
The $F$-terms which explicitly read
\bea
V_F &=& 40 d_{abc}d^{cde} ({H_1}^a {H_2}^b{H_1}_d{H_2}_e + {H_2}^a {H_3}^b{H_2}_d{H_3}_e + {H_1}^a {H_3}^b{H_1}_d{H_3}_e).
\eea
can be now written as
\bea
V_F &=& 40 {\rm tr} ({\hat N}^\dagger {\hat N}+{\hat M}^\dagger {\hat M}+{\hat L}^\dagger {\hat L}) .
\eea

\section{Supersymmetry and gauge symmetry breaking}
 
Consider the following vevs:
\be
M_0^{(1)}  = \begin{pmatrix} 
0 & 0 & 0 \cr 
0 & 0 & 0 \cr
0 & 0 & V
\end{pmatrix}, \hskip 1cm
M_0^{(2)}  = \begin{pmatrix} 
0 & 0 & 0 \cr 
0 & 0 & 0 \cr
V & 0 & 0
\end{pmatrix}\label{vev2}
\ee
for $H_3^{(1)}$ and $H_3^{(2)}$ respectively. These vevs leave the $SU(3)_c$ part of the gauge group unbroken
but trigger the spontaneous breaking of the rest.
More precisely, $M_0^{(1)}$ breaks the gauge group according to
\be
SU(3)_c\times SU(3)_L\times SU(3)_R \times U(1)_A \times U(1)_B \longrightarrow SU(3)_c\times SU(2)_L\times SU(2)_R\times U(1)
\ee
while $M_0^{(2)}$ according to
\be
SU(3)_c\times SU(3)_L\times SU(3)_R \times U(1)_A \times U(1)_B\longrightarrow  SU(3)_c\times SU(2)_L\times SU(2)_R'\times U(1)' \, .
\ee
The combination of the two gives \cite{Will}
\be
SU(3)_c\times SU(3)_L\times SU(3)_R  \times U(1)_A \times U(1)_B \longrightarrow SU(3)_c\times SU(2)_L\times U(1)_Y\, .
\ee
Electroweak (EW) breaking then proceeds by a second vev $v$, for example by \cite{MaZoup}
$M_{0}^{(1)} = {\rm diag}(v,v,V)$.
We first look at $V^{(1)}$ in the presence of the vevs.
Using the fact that the coefficients ${(G^A)}^b_a$ are antisymmetric in $a$ and $b$,
it is easy to see that for these vevs, the 
quadratic form $\langle  \phi | G^A | \phi \rangle$ vanishes identically in the vacuum, 
and so do the corresponding $SU(3)$ $D$-terms $D^A$.
The other terms give in the vacuum
\bea
V_{D_1} &=& 15 (V^2+2v^2)^2 \nonumber\\
V_{D_2} &=& \frac{5}{9} (V^2+2v^2-\theta_0^2)^2 \nonumber\\
V_F &=& \frac{40}{9} v^2(2 V^2 + v^2) \nonumber\\
V_{\rm soft} &=& \left(\frac{4R_2^2}{R_1^2R_3^2}-\frac{8}{R_2^2}\right) (V^2 + 2v^2) +
\left(\frac{4R_3^2}{R_1^2R_2^2}-\frac{8}{R_3^2}\right) (\theta^{(1)}_0)^2 \nonumber\\
&+&160\sqrt{3} 
\left(\frac{R_1}{R_2R_3}+\frac{R_2}{R_1R_3} + \frac{R_3}{R_1R_2}\right) V v^2 \, .
\eea 
As expected, already in the vacuum where EW symmetry is unbroken, supersymmetry is broken by both
$D$ and $F$-terms, in addition to its breaking by the soft terms.  
The potential is positive definite so we are looking for a 
vacuum solution with $V^{(1)}_0=0$.  
For simplicity we choose $R_1=R_2=R_3=R$
(strictly speaking in this case the manifold becomes nearly K\"ahler). Then, if the vevs satisfy the relation
\bea
(\theta^{(1)}_0)^2 &=& \frac{1}{10R^2} \Bigl[5R^2V^2 + 10R^2v^2 + 9 \nonumber\\
&+& \bigl( -675V^4R^4 -3100 V^2v^2 R^4 +270 V^2 R^2 -2900 v^4R^4 \nonumber\\
&+& 540 v^2R^2 
+27-21600\sqrt{3} Vv^2 R^3\bigr)^{1/2}
\Bigr] \, ,
\eea
the potential is zero at the minimum. We stress that in contrast to exactly supersymmetric theories, the 
zero of the potential at the minimum does not imply unbroken supersymmetry. This is because the
potential is a perfect square (which is a consequence of its higher dimensional origin from $F_{MN}F^{MN}$) with
the soft breaking terms included.  

It is interesting to notice that generically the solution makes sense when $V < 1/R$
and then if we set, with no loss of generality $V=1$, we find that the quantity under the square root is positive
if $v \sim {\cal O}( 0.1)$ for $R\sim {\cal O} (1/2)$. 
It is interesting that the desired hierarchy of scales is naturally generated by the 
structure of the scalar potential.

The analysis of $V^{(2)}$ in the presence of the second of the vevs in eq. (\ref{vev2}) is similar. The potential 
is zero at the minimum if the vev of $\theta^{(2)}$ satisfies
\be
(\theta^{(2)}_0)^2 = \frac{1}{10 R^2}\left(5V^2R^2 +9+
3\sqrt{-75V^4R^4+30V^2R^2+3}\right).
\ee
The vevs $\theta^{(1)}_0$ and $\theta^{(2)}_0$ need not be equal and $\theta^{(3)}_0$ can, but does not need 
to be zero.

\subsection{$U(1)$ structure and Yukawa couplings}

The breaking pattern of the bifundamental representations  that $V$ induces  is 
\bea
&&({\overline 3},1,3)_{(3,1/2)} \longrightarrow ({\overline 3},1,1+1+1)_{(3,1/2)}\label{rep1}\\
&&(3,{\overline 3},1)_{(-3,1/2)}\longrightarrow (3,2+1,1)_{(-3,1/2)}\label{rep2}\\
&&(1,3,{\overline 3})_{(0,-1)}\longrightarrow (1,2+1,1+1+1)_{(0,-1)}\label{rep3}
\eea
from which we can read off the representations under the SM gauge group and the extra
$U(1)$'s. From (\ref{rep1}) we obtain ${\bf \overline u}$, ${\bf \overline d}$ and ${\bf \overline D}$,
that is the two right handed quarks and an extra quark type state. From (\ref{rep2}) we obtain 
the quark doublet ${\bf Q}$ and the vector-like partner ${\bf D}$ of the extra quark.
Notice however that the extra quarks are not completely vector-like, since they have the same 
$U(1)_B$ charge. From (\ref{rep3}) we obtain the lepton doublet $L$, the right handed lepton
singlet ${\overline e}$, two right handed neutrinos and two electroweak doublets.
We will denote the latter doublets as $H_u$ and $H_d$ like in the minimal supersymmetric SM (MSSM). 
Notice that the scalar components of these doublets are the components of the $H_3$ Higgs field
that takes the vev $v$.
We will denote the former singlets as ${\overline N}_{1,2}$ 
while the singlet chiral superfields whose lowest component are the $\theta^{(l)}$
we will call ${\Theta}^{(l)}$.  
In the table we summarize the states contained in one family, with their $U(1)$ charges.
We have separated the MSSM spectrum from new states by a double line.
 
\vskip 0.2cm
\hskip 2cm
\begin{center}
\begin{tabular}{|c|c|c|c|}
\hline 
$ SU(3)_c\times SU(2)_L $ & $U(1)_Y$ & $ U(1)_A $ & $ U(1)_B $ \\
\hline \hline   
${\bf Q}\sim (3,2)$ & $1/6$ &$-3$ & $1/2$\\ \hline
${\bf \overline u}\sim ({\bar 3},1)$ & $-2/3$ & $3$ & $1/2$\\ \hline                                 
${\bf \overline d}\sim ({\bar 3},1) $ & $1/3$ & $3$ & $1/2$ \\ \hline
$ L \sim (1,2)$ & $-1/2$ & $0$ & $-1$ \\ \hline
${\overline e}\sim (1,1)$ & $1$ & $0$ & $-1$ \\ \hline
$H_u\sim (1,2)$ & $1/2$ & $0$ & $-1$ \\ \hline
$H_d\sim (1,2)$ & $-1/2$ & $0$ & $-1$ \\ \hline\hline
${\bf D}\sim (3,1)$ & $-1/3$ & $-3$ & $1/2$  \\ \hline
${\bf \overline D}\sim ({\bar 3},1)$ & $1/3$ & $3$ & $1/2$ \\ \hline
${\overline N_1}\sim (1,1)$ & $0$ & $0$ & $-1$ \\ \hline
${\overline N_2}\sim (1,1)$ & $0$ & $0$ & $-1$ \\ \hline
${\Theta}^{(1)}\sim (1,1)$ & $0$ & $0$ & $-1$ \\ \hline
\end{tabular}
\end{center}
\vskip 0.3cm
We immediately recognize that
\be
U(1)_A = -9 B\, ,
\ee
where $B$ is baryon number and $U(1)_B$ as a Peccei-Quinn type of symmetry.
Lepton number on the other hand does not appear to be a conserved symmetry (e.g. $L{\bf Q}{\bf \overline d}$ is allowed).
The presence of a conserved global baryon number is clearly a welcome feature from the point of view of
the stability of the proton.
The two extra $U(1)$'s at this stage, are both anomalous and at least one of them will remain anomalous after charge
redefinitions. 
They both break by the vev $V$, however their respective global subgroups remain at low energies and constrain
the allowed (non-renormalizable) operators in the superpotential.
Gauge invariance in the presence of the anomalous symmetries can be maintained by the addition of a specific combination of terms
to the low energy effective Lagrangian,
including a St\"uckelberg coupling and an axion-like interaction. 
These interactions introduce a new, phenomenologically interesting sector in the effective action \cite{cen}.
Let $A_\m$ be the anomalous $U(1)$ gauge field and $F_A$ its field strength. Then the terms that render the
action gauge invariant are
\be
{\cal L}_{\rm St-WZ} = \frac{1}{2}(\partial_\m a + M A_\m)^2 + c \frac{a}{M} F_A\wedge F_A + 
{\cal L}_{\rm an}\, .
\ee
The axion $a$ shifts under the anomalous symmetry so that the kinetic term is invariant
and coefficient $c$ is such that the Wess-Zumino term cancels the 1-loop anomaly ${\cal L}_{\rm an}$.
The scale $M$ is related to the vev $V$. 
These couplings are added by hand because they are not part of the 
the interactions of the original ten-dimensional gauge multiplet, neither can be generated by its
dimensional reduction. In fact, the axion field is the four-dimensional remnant of the two-form $B_{MN}$.
This is the (minimum) price to pay for neglecting the gravitational and two-form sectors 
(and actually also the second $E_8$ factor) along with the ten-dimensional anomaly cancellation mechanism,
for which their presence is essential \cite{GS}.

A few comments regarding the Yukawa sector are in order. Every operator originating from
the superpotential $d_{abc}H_1^aH_2^bH_3^c$ will appear at tree level.
At the quantum level operators that break the CSDR constraints and the supersymmetric structure will eventually 
develop, as long as they are gauge invariant. As an example of the former case,
the extra vector-like pair of quarks will develop a mass term in the $V$-vacuum
\be
\Theta^{(1)} {\bf \overline D}{\bf D}
\ee
which is a singlet. As an example of the latter, notice that in the quark sector
the standard Yukawa terms in the superpotential appear at tree level. 
In the lepton sector however the term $L{\overline e}H_d$ is not invariant under $U(1)_B$. 
An effective Yukawa coupling can come though from the higher-dimensional operator
\be
L{\overline e} H_d \left( \frac{{\theta^{(1) *}}}{M}\right)^3
\ee
in the $V$-vacuum, with $M$ a high scale such as the string scale and $\theta^{(1)*}$ the complex conjugate of $\theta^{(1)}$. 
Similar arguments apply to the entire lepton sector:
effective Yukawa couplings appear via higher dimensional operators
\be
L H_u {\overline N} \left( \frac{{\theta^{(1)*}}}{M}\right)^3\hskip 1cm
M  {\overline N} {\overline N}  \left( \frac{{\theta^{(1)*}}}{M}\right)^2\, .
\ee
Similar terms are generated for the second and third families.
Evidently, after electroweak symmetry breaking, fermion mass hierarchies and mixings can be generated \cite{FN}
not because the $U(1)$'s have  flavour dependent charges,
but from the different values that the vevs $\theta^{(l)}$ can have. 
A term that mixes flavours is, for example,
\be
L^{(1)}{\overline e}^{(2)} H_d^{(2)} \left( \frac{{\theta^{(1) *}}}{M}\right)\left( \frac{{\theta^{(2) *}}}{M}\right)^2\, ,
\ee
where we have made the flavour superscripts explicit on all fields.
A detailed global analysis of the model will be presented in a forthcoming work.

{\bf Acknowledgments.} We would like to thank I. Antoniadis, A. Chatzistavrakidis and D. L\"ust for discussions.
G.Z. is grateful to the Sommerfeld-LMU, MPI Munich and the ITP Heidelberg for warm hospitality.
This work was partially supported by the NTUA's basic research support program "PEVE" 2009 and 2010
and the European Union's ITN program "UNILHC" PITN-GA-2009-237920. 
 




\end{document}